# Analyzing the Instability of Large Language Models in Automated Bug Injection and Correction

Mehmet Bilal Er[1*], Nagehan İlhan[2], Umut Kuran[3]

[1,2,3]Department of Computer Engineering, Harran University, 63300 Şanlıurfa, Turkey

*Corresponding authors: bilal.er@harran.edu.tr

## Abstract

The use of Large Language Models (LLMs) in software engineering tasks is growing, especially in the areas of bug fixing and code generation. Nevertheless, these models often yield unstable results; when executed at different times with the same input, they can generate radically different code. The consistency of LLMs in bug-fixing tasks has not yet been thoroughly assessed, despite the fact that this instability has typically been discussed in the literature in relation to code generation. The purpose of this study is to look into how unstable an LLM like ChatGPT is when it comes to fixing code bugs. We examine the structural, syntactic, and functional variations among several fix recommendations made in response to the same prompt using code samples with various error types. Additionally, we assess how instability is affected by the temperature settings (0, 0.5, and 1) used for the model's deterministic operation. For a total of 20 problems in the experimental analysis, the model produced three fix suggestions at each temperature value, comparing nine distinct outputs for each problem. The Syntax Similarity and Output Equivalence Rate (OER) metrics were used to assess the outputs' structural and functional consistency. The results demonstrate that the model's outputs become much more unstable and variable as the temperature rises, with high temperatures showing especially high rates of functional failure. According to syntax similarity analyses, the suggested fixes show notable structural differences at high temperatures but are fairly similar at low temperatures. The purpose of this study is to provide important methodological insights into how LLM-based error correction systems can be applied more consistently in software development processes while also casting doubt on their dependability.

## 1. INTRODUCTION

LLMs have revolutionized software engineering in recent years with their exceptional performance in tasks such as code generation, error correction, and test scenario creation [1,2]. Models like OpenAI's ChatGPT can evaluate erroneous code provided in natural language and suggest correct, working versions [3]. This feature enables automation and accelerates software development processes. However, the generative power of LLMs also brings some fundamental issues. The most notable of these is uncertainty, which causes the same input (command) to produce markedly different outputs at different times [4]. The literature has partially examined how this affects code generation, but has not yet comprehensively examined how unstable LLMs are at error correction. Recent studies have shown that LLMs can generate code at near-competition levels using diversity and selection approaches. These studies also examine the increasingly mature research ecosystem of code-focused LLMs, the risks in evaluation processes, and open research questions. However, the issue of stability in LLM-based bug fixing processes has been relatively less addressed and insufficiently explored compared to the field of code generation [1,2]. The ability to automatically correct code errors is becoming more and more popular in both industry and academia [1,3]. However, it can seriously compromise software correctness, test reliability, and developer confidence to suggest a different fix every time an LLM is run against the same problematic piece of code [4,5]. Additionally, code quality, security, and maintenance expenses all depend on the functional equivalency (also known as semantic equivalency) of various suggested fixes [5,6]. The instability of LLMs in error correction tasks is examined empirically in this work from a variety of angles. The fixes produced by ChatGPT in various executions of the same prompt were examined using code fragments with various error types (such as syntactic errors, logic errors, null reference errors, etc.). QuixBugs is often used for benchmark-driven APR (Automatic Program Repair) studies, partly because it offers a combination of small defects and test oracles that can be executed, which is necessary to allow an objective comparison between works [7,8]. We ground this analysis on a controlled testbed, employing the QuixBugs benchmark, which pairs high-level buggy fragments with well-specified test oracles to experiment with multiple generations for a variety of defect types. In accordance with established practice for APR evaluation, we use multi-trial estimation and variance-aware reporting to prevent overfitting to single best-of-N findings. Test cases were used to evaluate the fixes' functional equivalence after they were compared both structurally and semantically. Furthermore, we used the temperature parameter to examine the degree to which the model resembled deterministic behavior. Reliability and repeatability should be taken into account, as the results show that LLMs can display a high degree of instability in error correction tasks. The goal of this research is to significantly advance methodologically the safer and more reliable application of LLMs in software engineering. Previous work suggests that non-determinism in code tasks decreases by but not disappears when temperature = 0 [4]; we therefore consider temperature (0.0, 0.5, 1.0) as an experimental variable and measure its impact on structural and semantic stability.The main contributions of this study are summarized as follows:

The major contributions of this study are summarized as follows:

- Stability-based evaluation design: For the same bug, we quantify the structural variability by character-level Levenshtein similarity and functional consistency by the test-based Output Equivalence Rate, OER, across successive generations under controlled temperature.
- Problem-level characterization on QuixBugs: Given a collection of twenty algorithmic defects, we characterize where instability is concentrated (certain classes of algorithms), and directly correlate measured variation with defect properties on a benchmark with known test oracles.
- Motivated by the evidence of sampling and selection at the core of modern code generation and repair, we derive stabilization techniques - multi-sample ensembling with test-aware selection; patch clustering with representative selection and timely standardization - to embed LLM repair as a probabilistic assistant.

The proposed stability-oriented LLM repair protocol can be used in a variety of software engineering environments. First, it can act as a CI (Continuous Integration) /CD (Continuous Delivery) admission gate, setting a multi-sample generation requirement and an OER threshold to filter out chance-passing patches before they get deployed. Second, it enables a repeatable verification step, which is especially important in safety-critical domains (e.g., finance, healthcare, automotive, embedded systems) where the risk of unstable fixes making it to production should be minimized. Third, for large codebases and monorepos, the ability to cluster structurally similar patches and select representatives can reduce maintenance overhead and code-review burden. Fourth, during migration and large-scale refactorings (API upgrade, language/runtime migration), distributional measures (e.g. Levenshtein dispersion, OER) can provide ongoing assurance of semantic preservation. Fifth, for model governance and comparative evaluation, the protocol provides objective stability metrics to benchmark LLMs and decoding settings and for research and education, it enables reproducibility using variance reporting and majority-vote test-aware acceptance.

The rest of this paper is organized in the following way: Section 2 discusses related work on code LLMs, LLM-based APR, and instability findings; Section 3 details the experimental setup (dataset, prompting, and metrics: Levenshtein and OER); Section 4 shows the stability results with respect to temperature and defect types; Section 5 makes comments on implications on reliability, reproducibility, and tooling; and Section 6 ends with limitations and future research directions.

## 2. RELATED WORK

A growing body of work shows that nondeterminism is a pervasive property of LLM-generated code rather than a corner case. In one of the most granular audits to date, ChatGPT produced bit-for-bit identical solutions to the same input in only 24.24% of CodeContests items, 49.00% of APPS, and 52.44% of HumanEval when run under default settings; all remaining trials diverged at the structural, syntactic, or functional level [9]. Subsequent studies echo this instability, noting that even specially adjusted models can shift back and forth between correct and incorrect behaviors across repeated runs on the very same task instance, particularly for more complex problems [10]. A recurrent explanation is sensitivity to seemingly minor small changes: slight edits to the input, duplication artifacts in the training set, or small input variations can push decoding toward different solution paths, producing output inconsistency [11–14]. Sampling hyperparameters adjust, but do not remove, this variability. Multiple reports find that lowering temperature reduces spread while leaving residual randomness intact [9,15]. In automatic bug-reproduction experiments, [16] identify a practical– diversity–alignment "sweet spot" around temperature ≈ 0.6, yet still observe significant run-to-run variation even at temperature = 0, attributable to other stochastic elements in the generation stack. Broader evaluations confirm that temperature and sampling choices induce noticeable performance oscillations, with higher temperatures widening the dispersion of pass rates and error modes [10,16]. Beyond aggregate accuracy, several works measure consistency along multiple similarity axes [9]. quantify semantic, syntactic, and structural similarity across repeated generations, documenting high difference on all three. Evidence from the broader NLP literature shows that pretrained models can either echo inputs or fail to generalize to perturbed contexts, which reduces cross-input consistency even for factual QA [11,17] for program-generation reliability. In software-engineering-specific audits, generated code frequently exhibits style violations, incorrect assertions, and functional faults, reinforcing that inconsistency spans formatting conventions, program logic, and observable behavior [13,14]. Reliability patterns are also strongly task-, model-, and context-dependent. Larger parameter counts or more elaborate inputing do not uniformly produce steadier outcomes; several controlled studies report task-contingent gains and regressions [18,19]. Small changes in model version or input phrasing can swing results substantially, complicating reproducibility and longitudinal comparisons [15,20]. Domain mismatch and data quality further matter: general-purpose models often underperform domain-tuned systems, and noisy/biased training corpora can propagate brittle behaviors into downstream coding tasks [16,21]. These findings align with broader surveys covering LLMs for software engineering, testing, and security, which document uneven reliability across pipelines, languages, and benchmarks—including vulnerability detection settings and large-code scenarios [22–26]. In response, the literature explores mitigation strategies that operate at generation time, through adaptation, or via downstream verification. Ensemble and collaborative schemes increase robustness by aggregating diverse drafts or model views: ensemble inputing has been shown to repair all tested hardware bugs in one security study, and self-collaboration improves pass rates by letting the model iteratively propose and refine candidates [27,28]. Targeted fine-tuning and RLHF can produce measurable gains in code-review-style tasks, improving both accuracy and comprehensibility [29], while hybrid workflows that pair LLMs with static analysis catch pragmatic bug patterns that generative decoding alone misses [30]. Complementary surveys catalog automatic correction approaches—agentic verify-and-repair loops, retrieval-augmented refinement, and feedback-driven editing—that systematically reduce, though do not abolish, nondeterminism-induced failures [31]. At the input design layer, evidence-based guidelines have been proposed to standardize procedures that improve run-to-run consistency and reduce evaluation variance [18]. Domain-specific lines of work—for instance HDL generation with VeriGen—further illustrate that specialized data and objectives can help, yet still exhibit sensitivity to task complexity and input idiosyncrasies [10]. Despite these advances, consensus remains that instability is not fully removed. Residual nondeterminism persists under low-temperature decoding, with ensembles, and after fine-tuning. External factors—API/version drift, training-data leakage, and limited transparency into corpora and filtering—continue to hinder faithful replication and long-horizon stability [20,21] . Adjacent evaluations on hallucination and reference accuracy reinforce the need for stronger guardrails and clearer reporting in safety-critical or high-stakes workflows [32,33]. Taken together, the literature motivates evaluation protocols that are themselves robust to variance: multi-seed/multi-input runs; semantic/syntactic/structural similarity scores; controlled sweeps over temperature and sampling; stratification by task type and domain specificity; and integration of automated verification. These practices produce more decision-relevant and reproducible estimates of reliability for code generation, bug reproduction, and code review—while making explicit the conditions under which current LLMs remain sensitive or unstable.

## 3. METHOD AND MATERIALS

This study examines the instability of LLMs in the software engineering context, specifically investigating their impact on code fixing (bug fixing) tasks. Inconsistency in model output is assessed by examining the structural, syntactic, and semantic variations in corrected code suggestions received at different times for the same faulty code snippet. In this context, an experimental analysis was conducted using the ChatGPT model, and the impact of model instability on the way code bugs are fixed was evaluated using multiple metrics. The study's core research questions are:
- Does ChatGPT produce consistent output across different runs for the same faulty code snippet?
- Is LLM instability more pronounced for certain types of bugs?
- Are different outputs functionally equivalent? Do the codes pass the same test?
- How does the LLM's temperature parameter affect the level of instability?

Based on these questions, both the output variability and the functionality of the output were measured.

### 3.1. Model Configuration and Prompt Design

The ChatGPT (GPT-4) model provided by OpenAI was used for code correction tasks. The model was tested with the following two different temperature parameters: Temperature = 1.0 (default): The model is expected to produce diversity in its output. Temperature = 0.0: The aim is to obtain more deterministic and repetitive answers. For each incorrect code segment, the model was queried nine consecutive times with the same prompt. The general prompt structure used was as follows: "The following code contains an error. Please fix the error and provide the correct working version:\n\npython\n<code>\n". The outputs produced from each trial were recorded separately and structured for analysis. The nine outputs obtained for each incorrect example were analyzed from two main perspectives: First, the corrected codes were compared in terms of syntactic similarity (Levenshtein distance). Second, Semantic Similarity (Semantic Similarity): It was examined whether the codes corrected with LLM models and then run produced the same output.

### 3.2. Levenshtein Distance for Syntactic Similarity

Vladimir Levenshtein, a Russian mathematician, introduced this metric in the context of error-correcting codes [34]. The Levenshtein Distance was initially used to quantify the difference between transmitted and received sequences in noisy communication channels. It later became widely adopted in computer science fields such as natural language processing, computational linguistics, and bioinformatics for comparing strings. Following that it started to be used to measure the syntactic dissimilarity between two sequences by counting the minimum number of single-character operations (insertions, deletions, or substitutions) required to change one string into the other [35–37]. Given two strings $s_1$ and $s_2$ of lengths $|s_1|$ and $|s_2|$, the Levenshtein Distance $D(i, j)$ between prefixes $s_1[1..i]$ and $s_2[1..j]$ is defined recursively as:

$$D(i,j) = \begin{cases} \max(i,j), & \text{if } \min(i,j) = 0 \\ \min \begin{cases} D(i-1,j) + 1 \\ D(i,j-1) + 1 \\ D(i-1,j-1) + c \end{cases}, & \text{otherwise} \end{cases} \quad (1)$$

To interpret this syntactic difference as a similarity score ranging from 0 (completely different) to 1 (identical), the normalized Levenshtein similarity is given by:

$$\text{Levenshtein Similarity}(s,t) = 1 - \frac{\text{Levenshtein }(s,t)}{\max(|s_1|,|s_2|)} \quad (2)$$

### 3.3 Output Equivalence Rate

The Output Equivalence Rate (OER) is a semantic similarity metric used to assess how often two programs or functions produce *equivalent outputs* for the same inputs. It is commonly used in software testing, automatic grading, and program repair validation. OER is considered an evaluation heuristic that quantifies how often two programs behave identically over a finite test set. It is frequently used in empirical software engineering and has become standard practice in evaluating equivalence in research papers [2,38,39]. Let $I = \{i_1, i_2, ..., i_n\}$ be a finite set of test inputs. For two programs P and Q, OER is defined as follows:

$$OER(P,Q) = \frac{|\{i \in I \mid P(i) = Q(i)\}|}{|I|} \quad (3)$$

That is, the OER is the ratio of input cases where both programs produce the same output, to the total number of test cases. OER = 1 indicates full output equivalence across the test set whereas OER = 0 indicates complete semantic divergence. In addition, OER values in (0,1) refer to partial equivalence. This measure provides insight into functional correctness, especially in contexts where syntactic similarity (e.g., Levenshtein Distance) itself is insufficient to judge program equivalence.

## 4. EXPERIMENTAL APPLICATIONS

## 4.1. Dataset

For use in experimental investigations, a comprehensive code sample pool comprising a diverse range of programming errors has been constructed. The primary objective of this pool is to provide a reliable and representative dataset for evaluating error detection and correction approaches. To achieve this, multiple sources were considered, with particular emphasis on the QuixBugs dataset, which has been widely adopted in software engineering research [40]. The QuixBugs dataset contains implementations of more than 40 distinct algorithmic problems, each available in both correct and faulty versions. These problems span a variety of fundamental algorithms, including search, sorting, numerical computations, and graph algorithms, making the dataset well-suited for benchmarking techniques related to automated program repair, bug injection, and bug correction. In the context of this study, we specifically focused on the Python implementations of QuixBugs, as they contain a rich collection of error-prone code samples that closely reflect real-world programming mistakes such as incorrect logic, boundary condition errors, and faulty control structures. By leveraging this dataset, we were able to ensure that our experiments were conducted on code samples that are both challenging and diverse, thereby improving the generalizability and robustness of the findings. Moreover, since QuixBugs is publicly available and widely recognized, its use also facilitates the reproducibility of our experiments and allows direct comparison with existing studies in the literature. Details of the QuixBugs dataset used in this study are provided in Table 1.

**Table 1.** Description of the QuixBugs Dataset Used in This Study

| Feature | Description |
| --- | --- |
| Purpose of use | Evaluation of automated program repair, bug injection, and bug correction methods |
| Number of algorithms | 20+ distinct algorithmic problems |
| Algorithm types | Search, sorting, numerical computations, graph algorithms |
| Versions | Each problem available in correct and faulty versions |
| Programming languages | Python and Java (this study focuses on Python versions) |
| Error types | Logic errors, boundary condition errors, faulty control structures |
| Characteristics | Rich variety of real-world programming mistakes; publicly available and widely used in the literature |
| Role in this research | Serves as the primary dataset for experimental analysis; ensures generalizability, robustness, and reproducibility |

## 4.2. Results analysis

The effect of LLMs on code remediation is methodically examined in this section. Every experiment was carried out using the popular and open-source QuixBugs dataset. This dataset includes both the right and wrong versions of 20 distinct algorithmic problems in different categories. The GPT-4 model, which could be accessed through the OpenAI API, was used for the experiments, and for every problem, nine examples were produced. Setting temperature = 0, 0.5, and 1.0 produced different outputs even though the model was guided by the same system prompts during each run. This made it possible for us to see how the LLM in code remediation is not deterministic.

The following metrics were computed in order to comprehend LLM's unpredictable behavior:

- Syntactic Similarity: Character-level metrics like Levenshtein distance are used to analyze the surface-level similarity between generated codes.
- Output Equivalence Rate (OER): This measures whether or not various solutions yield identical results.

Using the Levenshtein distance method, an analysis of syntactic similarity, or a measure of syntactic similarity, for a variety of programming problems with Temperature = 0 is presented in Table 2. Presenting different statistical measures of syntactic similarity between codes that correspond to specified problem names is the goal here. The table shows the standard deviation, variance, maximum and minimum similarity values, average similarity, and the proportion of low-similarity cases for each problem. The average Levenshtein similarity of the codes analyzed for each problem is first displayed in the "Average Similarity" column. The closer the value is to 1, the more similar the codes are to one another, since Levenshtein similarity is a metric that counts the character-based differences between two texts and is expressed as a similarity ratio. For instance, the "breadth_first_search" problem has an average similarity of 0.40, whereas the "reverse_linked_list_test" problem has an average similarity of up to 0.91. This shows that the breadth_first_search codes are very different from the reverse_linked_list_test codes, which are very similar. The distribution and variability of the codes' similarities are displayed in the "Variance" and "Standard Deviation" columns. While high variance and standard deviation values show notable differences between the codes, low values show that the codes are fairly similar. For instance, there are notable syntactical differences between the codes written for the "bitcount" problem, as evidenced by the variance of 0.076 and the standard deviation of 0.28. The highest and lowest similarity scores for the examples under examination are displayed in the "Maximum Similarity" and "Minimum Similarity" columns. As a result, certain codes for a particular problem might be extremely similar (for example, 1.00, which indicates perfect similarity), while others might be very different (for example, a value as low as 0.31). The "Low Similarity Ratio (<0.7)" column is among the table's most intriguing columns. The proportion of examples analyzed for each problem with a similarity value less than 0.7 is shown in this column. This ratio indicates how much the code created for the issue deviates from one another. The "bucketsort" problem, for instance, has a low similarity rate of 100%, indicating that all of the code is very different from one another; in contrast, the "reverse_linked_list_test" and "kth" problems have low similarity rates of 0%, indicating that the code is fairly similar. Overall, the table demonstrates that while some problems have more standard, formulaic, and similar solutions (e.g., reverse_linked_list_test, flatten, kth), others have a wide range of approaches (e.g., bucketsort, breadth_first_search, lis, sqrt). This could be caused by things like algorithm selections, solution diversity, and the complexity of the problem structure. In order to help us distinguish between problems with more varied and distinct approaches and those with more consistent and similar code, this table shows the degree of problem-specific diversity and standardization in terms of syntactic similarity. This offers significant insights in fields like error detection, similarity analysis, and code auto-fixing.

**Table 2**. Syntactic Similarity (Levenshtein) Table -Temperature =0

|   | Problem Name | Average Similarity | Variance | Maximum Similarity | Minimum Similarity | Standard Deviation | Low Similarity Ratio (<0.7) |
|---|---|---|---|---|---|---|---|
| 1 | reverse_linked_list_test | 0.91 | 0.01 | 1 | 0.78 | 0.1 | 0% |
| 2 | bucketsort | 0.63 | 0 | 0.65 | 0.6 | 0.02 | 100% |
| 3 | breadth_first_search | 0.4 | 0 | 0.4 | 0.39 | 0 | 100% |
| 4 | flatten | 0.99 | 0 | 0.99 | 0.99 | 0 | 0% |
| 5 | depth_first_search | 0.85 | 0 | 0.85 | 0.85 | 0 | 0% |
| 6 | shunting_yard | 0.8 | 0.016 | 1 | 0.6 | 0.13 | 17% |
| 7 | powerset | 0.62 | 0.002 | 0.65 | 0.53 | 0.05 | 100% |
| 8 | bitcount | 0.81 | 0.076 | 0.99 | 0.46 | 0.28 | 33% |
| 9 | detect_cycle_test | 0.78 | 0.027 | 0.92 | 0.54 | 0.17 | 50% |
| 10 | rpn_eval | 0.61 | 0 | 0.64 | 0.59 | 0.02 | 100% |
| 11 | next_permutation | 0.68 | 0 | 0.72 | 0.66 | 0.02 | 83% |
| 12 | to_base | 0.65 | 0 | 0.65 | 0.65 | 0 | 100% |
| 13 | shortest_path_lengths | 0.71 | 0.02 | 1 | 0.62 | 0.14 | 83% |
| 14 | subsequences | 0.86 | 0.003 | 0.98 | 0.84 | 0.06 | 0% |
| 15 | lcs_length | 0.85 | 0.005 | 0.94 | 0.78 | 0.07 | 0% |

|    | Problem Name | Average Similarity | Variance | Maximum Similarity | Minimum Similarity | Standard Deviation | Low Similarity Ratio (<0.7) |
|----|---|---|---|---|---|---|---|
| 16 | shortest_path_length_test | 0.67 | 0.078 | 1 | 0.31 | 0.28 | 33% |
| 17 | kth | 1 | 0 | 1 | 1 | 0 | 0% |
| 18 | lis | 0.65 | 0.002 | 0.68 | 0.55 | 0.05 | 100% |
| 19 | pascal | 0.84 | 0.028 | 0.95 | 0.63 | 0.17 | 33% |
| 20 | sqrt | 0.55 | 0.013 | 0.64 | 0.39 | 0.11 | 100% |

When the temperature parameter is set to 0.5, Table 3 displays the distribution of syntactic similarity (Levenshtein similarity) for various programming problems. The metrics in this table are identical to those in Table 1, but they enable us to examine the effects of raising the temperature on the structural variations and diversity of the code produced by the model. First, in artificial intelligence models, the temperature parameter is a hyperparameter that regulates the degree of diversity and randomness in outputs. The model tends to produce more consistent and similar solutions when the temperature drops, while producing more varied and distinct solutions as the temperature rises. Looking at the table, we can see that when the temperature is 0.5, the average similarity values typically drop. For instance, the average similarity dropped from 0.91 to 0.82 for the "reverse_linked_list_test" problem. This suggests that as the codes become less similar, diversity rises. Likewise, the similarity dropped from 0.80 to 0.79 for the "shunting_yard" problem and from 0.78 to 0.71 for the "detect_cycle_test" problem. The table's "Low Similarity Ratio (<0.7)" column shows a comparable rise. For instance, the "detect_cycle_test" problem's low similarity ratio rose from 50% to 67%, suggesting a rise in the percentage of examples with similarity below 0.7 and a wider range of solutions. In contrast, the "pascal" problem demonstrated a significant rise in this table, dropping from 33% to 0%, suggesting that the solutions became more uniform and comparable. The "flatten," "kth," and "depth_first_search" problems, for instance, showed high similarity values across nearly all metrics, while other problems continued to show low diversity and high similarity. This suggests that the solution paths are comparatively standard and stable since the solutions to these issues are structurally independent of variations in the model temperature. A greater range of similarity rates among various solutions and an increase in diversity are supported by the variance and standard deviation values, which typically show slight increases. The standard deviation, for instance, rose from 0.14 to 0.21 for the "shortest_path_lengths" problem and from 0.28 to 0.31 for the "bitcount" problem. In conclusion, the model's code generation becomes more diverse at a temperature of 0.5, which results in a rise in low similarity rates and a decrease in syntactic similarity between codes. Nevertheless, some problems retain high similarity even when the temperature is raised, especially those that involve standard and distinct algorithmic structures. Understanding how model outputs change with temperature and how syntactic similarities change based on the type of problem is made easier with the help of this table.

**Table 3.** Syntactic Similarity (Levenshtein) Table - Temperature =0.5

|    | Problem Name | Average Similarity | Variance | Maximum Similarity | Minimum Similarity | Standard Deviation | Low Similarity Ratio (<0.7) |
|----|---|---|---|---|---|---|---|
| 1  | reverse_linked_list_test | 0.82 | 0.002 | 0.87 | 0.78 | 0.05 | 0% |
| 2  | bucketsort | 0.63 | 0.001 | 0.65 | 0.61 | 0.02 | 100% |
| 3  | breadth_first_search | 0.4 | 0 | 0.4 | 0.39 | 0 | 100% |
| 4  | flatten | 0.99 | 0 | 0.99 | 0.99 | 0 | 0% |
| 5  | depth_first_search | 0.85 | 0 | 0.85 | 0.85 | 0 | 0% |
| 6  | shunting_yard | 0.79 | 0.001 | 0.83 | 0.78 | 0.03 | 0% |
| 7  | powerset | 0.63 | 0.001 | 0.65 | 0.6 | 0.02 | 100% |
| 8  | bitcount | 0.81 | 0.095 | 0.99 | 0.46 | 0.31 | 33% |
| 9  | detect_cycle_test | 0.71 | 0.038 | 0.92 | 0.54 | 0.19 | 67% |
| 10 | rpn_eval | 0.62 | 0 | 0.64 | 0.6 | 0.02 | 100% |
| 11 | next_permutation | 0.68 | 0.001 | 0.72 | 0.66 | 0.03 | 67% |

| | | | | | | |
|---|---|---|---|---|---|---|
| 12 | to_base | 0.65 | 0 | 0.65 | 0.65 | 0 | 100% |
| 13 | shortest_path_lengths | 0.76 | 0.043 | 1 | 0.62 | 0.21 | 67% |
| 14 | subsequences | 0.89 | 0.006 | 0.98 | 0.84 | 0.08 | 0% |
| 15 | lcs_length | 0.86 | 0.005 | 0.94 | 0.79 | 0.07 | 0% |
| 16 | shortest_path_length_test | 0.68 | 0.108 | 1 | 0.34 | 0.33 | 33% |
| 17 | kth | 1 | 0 | 1 | 1 | 0 | 0% |
| 18 | lis | 0.63 | 0.005 | 0.68 | 0.55 | 0.07 | 100% |
| 19 | pascal | 0.95 | 0 | 0.95 | 0.95 | 0 | 0% |
| 20 | sqrt | 0.53 | 0.011 | 0.62 | 0.42 | 0.11 | 100% |

When the temperature parameter is set to 1, Table 4 displays syntactic similarity (Levenshtein similarity) measurements for various programming problems. Compared to the first two tables, this one shows an even greater diversity and variation in the model's output. Greater syntactic variation between the codes results from the model's ability to generate high levels of randomness and diversity in its responses when the temperature parameter is set to 1. Each problem's average similarity, variance, maximum and minimum similarity, standard deviation, and low similarity ratio (<0.7) are displayed in the table. Compared to the earlier tables, the average similarity values are lower. The average similarity for "reverse_linked_list_test," for instance, is 73%, which is less than the earlier values of 0.91 (temp=0) and 0.82 (temp=0.5). This indicates that the model produces more diverse and distinct code as the temperature increases. For the "breadth_first_search" problem, the average similarity is only 0.45, meaning the solutions are quite different. For This suggests that as the temperature rises, the model generates more varied and unique code. With an average similarity of just 0.45 for the "breadth_first_search" problem, the solutions are significantly dissimilar. The average similarity stays low, ranging from 0.59 to 0.65, for problems like "bucketsort" and "powerset." Minimum similarity values are significantly lower (e.g., 0.41 or 0.33), indicating significant differences between the codes, while maximum similarity values (e.g., 0.96 for "reverse_linked_list_test") are still high, indicating that some examples are fairly similar. In comparison to the other tables, the variance and standard deviation values have generally increased, indicating a broad range of code similarity values and boosting diversity. For many problems, the "Low Similarity Rate (<0.7)" column displays relatively high rates: Examples of problems with a low similarity rate of 100% include "bucketsort," "powerset," "rpn_eval," "next_permutation," "to_base," "lis," and "sqrt." This suggests that every generated code differs significantly from the others in terms of syntax. However, issues such as "flatten," "kth," and "pascal" continue to have high similarity rates (0% low similarity rate), indicating that code diversity is still restricted and that solutions are still more standardized. The model produces a far greater range of codes when the temperature value is raised to 1, which results in a drop in average syntactic similarity and an increase in low similarity rates. Some problems still have standard solutions, even though the distribution of code similarity is more variable. The effect of the temperature parameter on AI model code generation is evident in this table, which indicates that the model generates more varied and imaginative but less reliable codes at high temperatures.

**Table 4.** Syntactic Similarity (Levenshtein) Table - Temperature =1

| | Problem Name | Average Similarity | Variance | Maximum Similarity | Minimum Similarity | Standard Deviation | Low Similarity Ratio (<0.7) |
|---|---|---|---|---|---|---|---|
| 1 | reverse_linked_list_test | 0.73 | 0.083 | 0.96 | 0.41 | 0.29 | 33% |
| 2 | bucketsort | 0.65 | 0 | 0.65 | 0.65 | 0 | 100% |
| 3 | breadth_first_search | 0.45 | 0.002 | 0.48 | 0.4 | 0.04 | 100% |
| 4 | flatten | 0.99 | 0 | 0.99 | 0.99 | 0 | 0% |
| 5 | depth_first_search | 0.9 | 0.002 | 0.94 | 0.85 | 0.04 | 0% |
| 6 | shunting_yard | 0.8 | 0.029 | 0.99 | 0.66 | 0.17 | 33% |
| 7 | powerset | 0.59 | 0.008 | 0.65 | 0.49 | 0.09 | 100% |
| 8 | bitcount | 0.81 | 0.095 | 0.99 | 0.46 | 0.31 | 33% |

| | | | | | | |
|---|---|---|---|---|---|---|
| 9 | detect_cycle_test | 0.79 | 0.033 | 0.92 | 0.58 | 0.18 | 33% |
| 10 | rpn_eval | 0.6 | 0.002 | 0.63 | 0.55 | 0.04 | 100% |
| 11 | next_permutation | 0.63 | 0.007 | 0.69 | 0.53 | 0.09 | 100% |
| 12 | to_base | 0.65 | 0 | 0.65 | 0.65 | 0 | 100% |
| 13 | shortest_path_lengths | 0.64 | 0.006 | 0.71 | 0.56 | 0.07 | 67% |
| 14 | subsequences | 0.84 | 0 | 0.84 | 0.83 | 0.01 | 0% |
| 15 | lcs_length | 0.67 | 0.097 | 0.94 | 0.33 | 0.31 | 33% |
| 16 | shortest_path_length_test | 0.67 | 0.082 | 0.84 | 0.34 | 0.29 | 33% |
| 17 | kth | 1 | 0 | 1 | 1 | 0 | 0% |
| 18 | lis | 0.67 | 0 | 0.68 | 0.66 | 0.01 | 100% |
| 19 | pascal | 0.95 | 0 | 0.95 | 0.95 | 0 | 0% |
| 20 | sqrt | 0.64 | 0.001 | 0.65 | 0.61 | 0.02 | 100% |

Box plots illustrating syntactic similarity (Levenshtein) distributions across problems for three distinct temperature values (0, 0.5, and 1) are displayed in Figure 1. This plot shows how similar the model-generated solutions are to the right code at each temperature level, as well as how these similarities are distributed. At temperature 0, similarities are high and constant; at temperature 0.5, there are minor variations; and at temperature 1, diversity rises and similarity sharply declines.

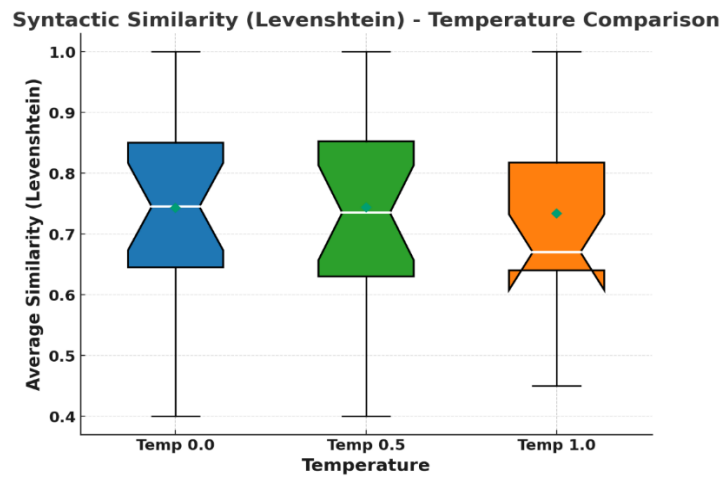

**Figure 1.** Syntactic Similarity (Levenshtein) – Temperature Comparison

As shown in Figure 2, with higher sampling temperature, syntactic alignment (mean Levenshtein similarity) tends to decrease, and cross-task dispersion tends to increase. Canonical, template-like issues, including flatten and kth, are ceiling-insensitive in all settings (~0.99–1.00), meaning that they are temperature-insensitive. In comparison, the problems with multiple realizations are more easily degraded: reverse-linked-list-test decreases between 0.91 and 0.82 to 0.73, next-permutation between 0.68 and 0.68 to 0.63, and lcs-length between 0.85 and 0.86 to 0.67.

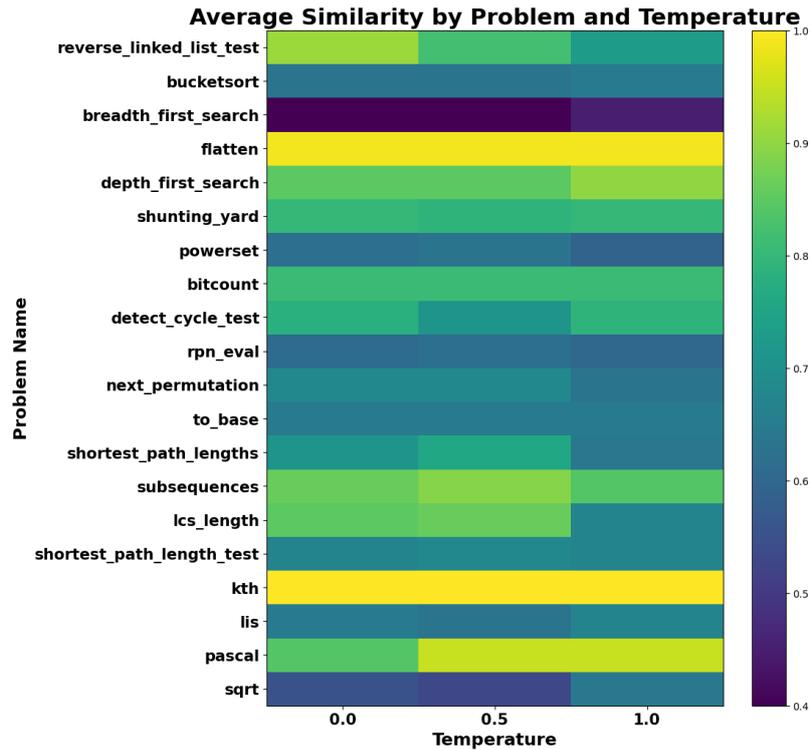

**Figure 2.** Average Similarity by Problem and Tempreature

There are also task nuances in Figure 2. Other graph/traversal problems do not work this way: shortest_path_lengths gains at T=0.5 (0.71 0.76), but loses at T=1.0 (0.64), depth_first_search gains slightly at T=1.0 (0.85 0.90). Some tasks are moderately stochastic; pascal increases between 0.84 and 0.95 at T=0.5-1.0. Otherwise they are quite stable irrespective of temperature (e.g. to_base 0.65; rpn_eval 0.6062). The performance results of the tests conducted with the Temperature=0.0 setting are detailed in Table 5, and the overall OER (Mean Equivalent Output Ratio) value was determined to be 0.7000. The number of successes and failures, statistical measurements (mean, variance, and standard deviation), and success percentages are displayed for each of the three trials that were conducted on 20 distinct problems. The table shows that the following problems had 100% success in every trial and were classified as "Fully Successful": bitcount, breadth_first_search, depth_first_search, flatten, kth, lcs_length, lis, next_permutation, shortest_path_lengths, subsequences, and to_base. This demonstrates that the model performs consistently and without errors in these kinds of problems. However, the rpn_eval and shortest_path_length_test problems were recorded as 0%, putting them in the "Failed" category. In contrast, the success rate for the bucketsort, pascal, powerset, reverse_linked_list_test, and shunting_yard problems is 33.33%. These findings show that the model has serious issues with these kinds of problems. The sqrt and detect_cycle_test problems in the intermediate region received a success rate of 66.67%, meaning that even though some inputs were correct, consistent success was not attained across all tests. Overall, the table makes it abundantly evident that although the model did very well in some algorithm types at Temperature = 0.0, sorting, combination generation, and specific data structure operations were its weakest points. This makes it evident which problem types the model excels at in deterministic mode and where it needs to be improved.

.

**Table 5.** Temperature=0.0 general OER=0.7000

| No | Problem | Count | Successes | Failures | Mean | Variance | StdDev | Success Rate (%) | Category |
|---|---|---|---|---|---|---|---|---|---|
| 1 | bitcount | 3 | 3 | 0 | 1.00 | 0.00 | 0.00 | 100.00 | Fully Successful |
| 2 | breadth_first_search | 3 | 3 | 0 | 1.00 | 0.00 | 0.00 | 100.00 | Fully Successful |
| 3 | bucketsort | 3 | 1 | 2 | 0.33 | 0.33 | 0.58 | 33.33 | Failed |
| 4 | depth_first_search | 3 | 3 | 0 | 1.00 | 0.00 | 0.00 | 100.00 | Fully Successful |
| 5 | detect_cycle_test | 3 | 2 | 1 | 0.67 | 0.33 | 0.58 | 66.67 | Partially Successful |
| 6 | flatten | 3 | 3 | 0 | 1.00 | 0.00 | 0.00 | 100.00 | Fully Successful |
| 7 | kth | 3 | 3 | 0 | 1.00 | 0.00 | 0.00 | 100.00 | Fully Successful |
| 8 | lcs_length | 3 | 3 | 0 | 1.00 | 0.00 | 0.00 | 100.00 | Fully Successful |
| 9 | lis | 3 | 3 | 0 | 1.00 | 0.00 | 0.00 | 100.00 | Fully Successful |
| 10 | next_permutation | 3 | 3 | 0 | 1.00 | 0.00 | 0.00 | 100.00 | Fully Successful |
| 11 | pascal | 3 | 1 | 2 | 0.33 | 0.33 | 0.58 | 33.33 | Failed |
| 12 | powerset | 3 | 1 | 2 | 0.33 | 0.33 | 0.58 | 33.33 | Failed |
| 13 | reverse_linked_list_test | 3 | 1 | 2 | 0.33 | 0.33 | 0.58 | 33.33 | Failed |
| 14 | rpn_eval | 3 | 0 | 3 | 0.00 | 0.00 | 0.00 | 0.00 | Failed |
| 15 | shortest_path_length_test | 3 | 0 | 3 | 0.00 | 0.00 | 0.00 | 0.00 | Failed |
| 16 | shortest_path_lengths | 3 | 3 | 0 | 1.00 | 0.00 | 0.00 | 100.00 | Fully Successful |
| 17 | shunting_yard | 3 | 1 | 2 | 0.33 | 0.33 | 0.58 | 33.33 | Failed |
| 18 | sqrt | 3 | 2 | 1 | 0.67 | 0.33 | 0.58 | 66.67 | Partially Successful |
| 19 | subsequences | 3 | 3 | 0 | 1.00 | 0.00 | 0.00 | 100.00 | Fully Successful |
| 20 | to_base | 3 | 3 | 0 | 1.00 | 0.00 | 0.00 | 100.00 | Fully Successful |

The model's performance on a range of problems at a temperature of 0.5 is shown in Table 6. At this temperature, the average equivalent output ratio (OER) was determined to be 66.67% overall. The success and failure rates derived from conducting three trials each on a total of twenty distinct problems are shown in the table. The effectiveness of the model at this temperature setting for each type of problem is shown by the success rates and other statistical data. The issues in the "Fully Successful" category are the first notable group in the table. Bitcount, breadth_first_search, depth_first_search, flatten, kth, lcs_length, lis, pascal, subsequences, and to_base are some of these issues. The model achieved a 100% success rate for these problems, producing output that was accurate and consistent throughout all three trials. This indicates that for specific algorithm and data structure types, the model offers high reliability even at this temperature setting. The "Partially Successful" category includes another group. This group includes problems like sqrt, shortest_path_lengths, powerset, and next_permutation. The model's success rate on these problems was roughly 66.7%, which means that while it failed in some examples, it produced accurate results in others. This illustrates how unstable the model is for these kinds of problems and how, in certain cases, the likelihood of error rises with temperature. The "Failed" group, which comprises problems like bucketsort, detect_cycle_test, reverse_linked_list_test, rpn_eval, shortest_path_length_test, and shunting_yard, is another especially notable category. Detect_cycle_test and reverse_linked_list_test are two of these issues that stand out as ones where the model failed all three times and failed to generate any accurate output at all. This implies that the model has serious issues with these kinds of issues and that the current temperature parameter needs to be improved in these areas. Overall, Table 5 shows that setting the temperature to 0.5 causes a significant drop in the model's performance on some problem sets. On certain problems, the model's output becomes more random, which has a detrimental effect on consistency and stability. For many problems, however, comparatively high accuracy rates are still maintained. These findings show which problem types need performance enhancements depending on the temperature parameter and offer important insight into how various temperature settings affect model behavior.

**Table 6.** Temperature=0.5 general OER=0.6667

| No | Problem | Count | Successes | Failures | Mean | Variance | StdDev | Success Rate (%) | Category |
|----|---------|-------|-----------|----------|------|----------|--------|------------------|----------|
| 1 | bitcount | 3 | 3 | 0 | 1.00 | 0.00 | 0.00 | 100.00 | Fully Successful |
| 2 | breadth_first_search | 3 | 3 | 0 | 1.00 | 0.00 | 0.00 | 100.00 | Fully Successful |
| 3 | bucketsort | 3 | 1 | 2 | 0.33 | 0.33 | 0.58 | 33.33 | Failed |
| 4 | depth_first_search | 3 | 3 | 0 | 1.00 | 0.00 | 0.00 | 100.00 | Fully Successful |
| 5 | detect_cycle_test | 3 | 0 | 3 | 0.00 | 0.00 | 0.00 | 0.00 | Failed |
| 6 | flatten | 3 | 3 | 0 | 1.00 | 0.00 | 0.00 | 100.00 | Fully Successful |
| 7 | kth | 3 | 3 | 0 | 1.00 | 0.00 | 0.00 | 100.00 | Fully Successful |
| 8 | lcs_length | 3 | 3 | 0 | 1.00 | 0.00 | 0.00 | 100.00 | Fully Successful |
| 9 | lis | 3 | 3 | 0 | 1.00 | 0.00 | 0.00 | 100.00 | Fully Successful |
| 10 | next_permutation | 3 | 2 | 1 | 0.67 | 0.33 | 0.58 | 66.67 | Partially Successful |
| 11 | pascal | 3 | 3 | 0 | 1.00 | 0.00 | 0.00 | 100.00 | Fully Successful |
| 12 | powerset | 3 | 2 | 1 | 0.67 | 0.33 | 0.58 | 66.67 | Partially Successful |
| 13 | reverse_linked_list_test | 3 | 0 | 3 | 0.00 | 0.00 | 0.00 | 0.00 | Failed |
| 14 | rpn_eval | 3 | 1 | 2 | 0.33 | 0.33 | 0.58 | 33.33 | Failed |
| 15 | shortest_path_length_test | 3 | 0 | 3 | 0.00 | 0.00 | 0.00 | 0.00 | Failed |
| 16 | shortest_path_lengths | 3 | 2 | 1 | 0.67 | 0.33 | 0.58 | 66.67 | Partially Successful |
| 17 | shunting_yard | 3 | 0 | 3 | 0.00 | 0.00 | 0.00 | 0.00 | Failed |
| 18 | sqrt | 3 | 2 | 1 | 0.67 | 0.33 | 0.58 | 66.67 | Partially Successful |
| 19 | subsequences | 3 | 3 | 0 | 1.00 | 0.00 | 0.00 | 100.00 | Fully Successful |
| 20 | to_base | 3 | 3 | 0 | 1.00 | 0.00 | 0.00 | 100.00 | Fully Successful |

When the temperature parameter is set to 1.0, Table 7 shows how well the model performs on a variety of problems. The model's accuracy rate slightly drops at this temperature, as indicated by the overall average equivalent output rate (OER), which is roughly 61.67%. For every problem, three trials were carried out; the table displays the number of successes and failures as well as statistical measurements. Bitcount, bucketsort, depth_first_search, flatten, kth, pascal, subsequences, and to_base are among the problems that were fully solved. The model achieved a 100% success rate in all three trials on these problems. This illustrates that in certain problem types, the model can continue to function with high consistency even as the temperature rises. On the other hand, the model faced difficulties and typically achieved low success rates when faced with problems like breadth_first_search, detect_cycle_test, next_permutation, powerset, reverse_linked_list_test, rpn_eval, shortest_path_length_test, shortest_path_lengths, and shunting_yard. For instance, the model's performance on the rpn_eval and shortest_path_length_test problems was 0% across all trials. This suggests that at temperature 1.0, the model's output became more random and fluctuated, which reduced its stability. Problems like lcs_length, lis, and sqrt fall under the partial success category. These problems had success rates of about 66.7%, meaning that the model was right in some cases but wrong in others. In comparison to the earlier temperature values, the model's overall stability and success rate declined at temperature 1.0. Inconsistency on some problems resulted from the increased randomness in the model outputs caused by the higher temperature. On some problem types, the model continued to perform well, though. This table gives guidance on which issues need to be fixed and clearly illustrates how temperature adjustment affects model performance.

**Table 7.** Temperature=1 genel OER=0.6167

| No | Problem | Count | Successes | Failures | Mean | Variance | StdDev | Success Rate (%) | Category |
|---|---|---|---|---|---|---|---|---|---|
| 1 | bitcount | 3 | 3 | 0 | 1.00 | 0.00 | 0.00 | 100.00 | Fully Successful |
| 2 | breadth_first_search | 3 | 1 | 2 | 0.33 | 0.33 | 0.58 | 33.33 | Failed |
| 3 | bucketsort | 3 | 3 | 0 | 1.00 | 0.00 | 0.00 | 100.00 | Fully Successful |
| 4 | depth_first_search | 3 | 3 | 0 | 1.00 | 0.00 | 0.00 | 100.00 | Fully Successful |
| 5 | detect_cycle_test | 3 | 1 | 2 | 0.33 | 0.33 | 0.58 | 33.33 | Failed |
| 6 | flatten | 3 | 3 | 0 | 1.00 | 0.00 | 0.00 | 100.00 | Fully Successful |
| 7 | kth | 3 | 3 | 0 | 1.00 | 0.00 | 0.00 | 100.00 | Fully Successful |
| 8 | lcs_length | 3 | 2 | 1 | 0.67 | 0.33 | 0.58 | 66.67 | Partially Successful |
| 9 | lis | 3 | 2 | 1 | 0.67 | 0.33 | 0.58 | 66.67 | Partially Successful |
| 10 | next_permutation | 3 | 1 | 2 | 0.33 | 0.33 | 0.58 | 33.33 | Failed |
| 11 | pascal | 3 | 3 | 0 | 1.00 | 0.00 | 0.00 | 100.00 | Fully Successful |
| 12 | powerset | 3 | 1 | 2 | 0.33 | 0.33 | 0.58 | 33.33 | Failed |
| 13 | reverse_linked_list_test | 3 | 1 | 2 | 0.33 | 0.33 | 0.58 | 33.33 | Failed |
| 14 | rpn_eval | 3 | 0 | 3 | 0.00 | 0.00 | 0.00 | 0.00 | Failed |
| 15 | shortest_path_length_test | 3 | 0 | 3 | 0.00 | 0.00 | 0.00 | 0.00 | Failed |
| 16 | shortest_path_lengths | 3 | 1 | 2 | 0.33 | 0.33 | 0.58 | 33.33 | Failed |
| 17 | shunting_yard | 3 | 1 | 2 | 0.33 | 0.33 | 0.58 | 33.33 | Failed |
| 18 | sqrt | 3 | 2 | 1 | 0.67 | 0.33 | 0.58 | 66.67 | Partially Successful |
| 19 | subsequences | 3 | 3 | 0 | 1.00 | 0.00 | 0.00 | 100.00 | Fully Successful |
| 20 | to_base | 3 | 3 | 0 | 1.00 | 0.00 | 0.00 | 100.00 | Fully Successful |

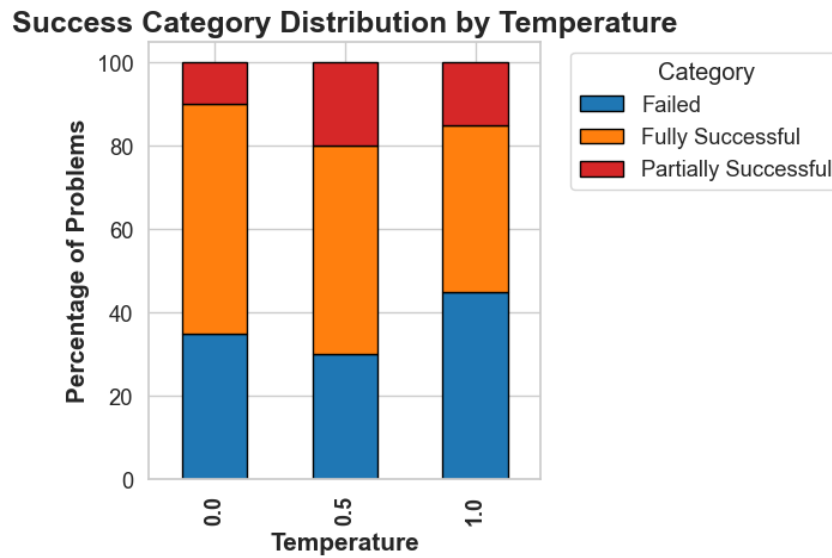

**Figure 3.** Success Category Distribution by Temperature

Figure 3 indicates the change in outcome composition with increase in temperature. At T=0.0, tasks are predominantly Fully Successful (55%), Failed (35%) and Partially successful (10%). At T=0.5, the share-fully-successful declines to 50 and the partial results increase to 20 (failures decrease slightly to 30), so there is more variability with no catastrophic collapse. At T=1.0, the decline in reliability is obvious: Fully Successful and Failed are 40% and 45% (partials 15%). This monotonic decrease in fully-correct solutions (55% to 50% to 40%), and the net increase in failures (35% to 30% to 45%), reflect the overall OER decrease in Tables 4, 5, 6 (0.70 to 0.67 to 0.62).

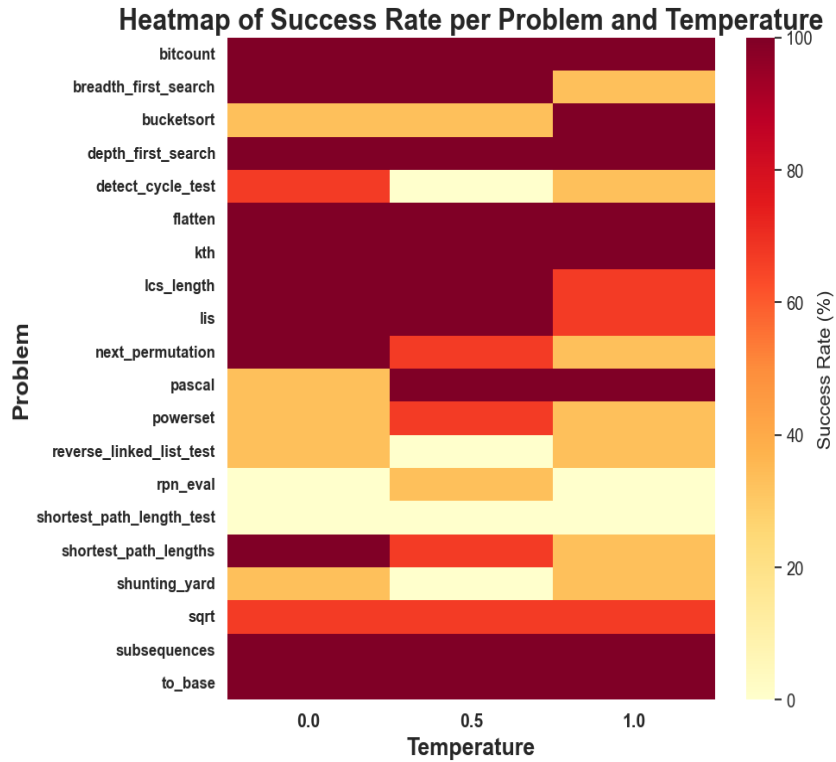

**Figure 4.** Heatmap of Success Rate per Problem and Temperature

Figure 4 is a heatmap of success rates of the 20 benchmark problems when varying temperature settings are applied. The findings indicate a response to randomness which is heterogeneous: some of the problems, including bitcount, flatten, kth, lcs length, subsequences, and to base, are consistently reliable with near-perfect accuracy in all settings, whereas others, including rpn eval and shortest path length test, are almost completely ineffective, regardless of temperature. Some of the functions such as breadth basis search, detect cycle test, lis, next permutation and shortest path lengths are very sensitive because they work fine at T=0.0 but decay very quickly as temperature increases. Smaller classes, such as bucketsort, pascal and powerset, display non-monotonic behaviour, with the performance increasing at intermediate temperatures and then decreasing again at T=1.0, indicating that a small dose of stochasticity may sometimes counteract brittle deterministic errors. Combined, Figure 4 reveals the disproportionate effect of temperature change on problem type, with deterministic and arithmetic tasks largely resistant to change, but graph traversal and logical reasoning problems plummeting in quality as randomness increases.

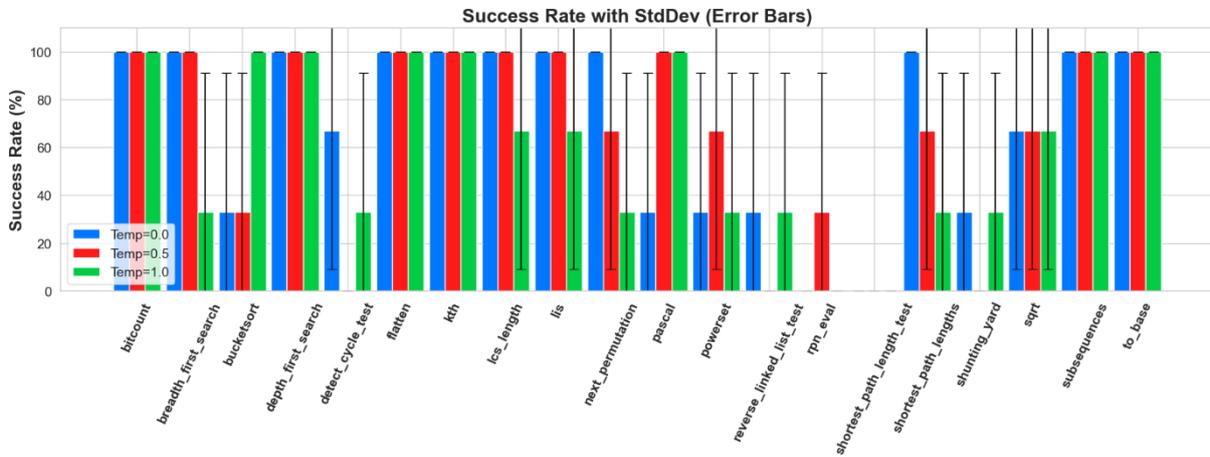

**Figure 5.** Success Rate with StdDev

Figure 5 shows the average success per problem with error bars of standard deviation values in the three temperatures. Not only the mean performance of each problem, but also the variability of outcomes are visible in the grouped bar chart. Bitcount, flatten, kth, subsequences, and to base assignments have high success rates with very small variance, and the results support the sensitivity of these assignments to temperature. Conversely, other issues such as rpn_eval and shortest-path-length-test are at or close to zero in all environments, indicating that there are systematic failures in spite of decoding strategy. Some tasks such as breadth_first_search, lis, next perm, and shunting yard show large decreases in success as the temperature goes up, and their error bars are very large, reflecting high instability over different runs. The interesting thing is that pascal and powerset perform better at intermediate temperature levels, with high variance, implying that stochasticity can sometimes benefit the model and reliability can be sacrificed.

## 5. DISCUSSION

The results systematically demonstrate the instability of LLM models in error correction tasks. Findings indicate that as the temperature parameter increases, both structural and functional consistency in the model's outputs decrease. While relatively more stable results are obtained at a temperature of 0.0, diversity increases at values of 0.5 and 1.0, but functional success decreases. The decline in the average OER value from 0.70 to 0.62 clearly confirms this trend. The results also vary depending on the type of problem. The model operates with high accuracy and low variance in simple, template-based, or math-based tasks (such as flatten, kth, bitcount), while stability rapidly deteriorates and success rates decline, particularly in graph-based and control flow-intensive tasks (such as breadth_first_search, detect_cycle_test, shortest_path_length_test). This situation reveals that the model has strong representations in some algorithmic structures, but its sensitivity increases in tasks requiring logical reasoning and state tracking. Another important finding is that even when the temperature is set to zero, fully deterministic behavior cannot be achieved. The variations observed in Levenshtein similarity values and the failures recorded for certain problems indicate that LLMs depend not only on the temperature parameter but also on sampling methods and systemic factors. Therefore, it is concluded that the expectation of determinism is limited and that the "temperature=0" approach does not provide a precise solution. Methodologically, examining structural similarity and functional equivalence together provides an important contribution. Relying solely on superficial similarity metrics could have overlooked functional failures. This study highlights the necessity of test-based measurements and the need to consider semantic equivalence in reliability assessments.

## 6. CONCLUSION

The purpose of this study was to systematically examine the performance of LLMs in error correction tasks. Twenty different problems selected from the QuixBugs dataset were used in the experimental analysis, and the ChatGPT (GPT-4) model was run nine times for each problem at three different temperature values (0.0, 0.5, 1.0). The total of 540 outputs obtained were evaluated in terms of both structural similarity (syntactic similarity measured by Levenshtein distance) and functional equivalence (Output Equivalence Rate, OER). This allowed the diversity and consistency among the different solutions produced for the same error to be quantitatively revealed. The results show that as the temperature value increases, the model produces more diverse but less reliable outputs. The decrease in the average OER value from 0.70 to 0.62 proves that functional success decreases with temperature. However, the findings are not homogeneous for all tasks. Problems with template-like properties and deterministic solutions (flatten, kth, bitcount) are solved with high success at all temperatures, while graph-based or control flow-intensive problems (breadth_first_search, detect_cycle_test, shortest_path_length_test) stand out with high failure rates. Notably, even at a temperature of 0.0, full determinism could not be achieved, indicating that the randomness of LLMs cannot be explained solely by the temperature parameter. These findings indicate that LLMs can be a powerful support tool for error correction tasks, but reliability issues pose a significant limiting factor. One-off and unsupervised use carries risks in terms of software accuracy and security. Therefore, methods such as multiple sampling, test-based selection, determining OER threshold values, and standardizing patches through clustering will enable LLM-based error correction systems to be integrated more reliably into software development processes.

Future work should extend this research to larger-scale and multi-file projects, different programming languages, and different LLM architectures; it should also be supported by more advanced analyses, such as AST-based or semantic similarity metrics. Long-term stability testing and examining the effects of model version changes also emerge as important research areas. In conclusion, LLMs are potentially useful tools for error correction; however, their instability should not be overlooked, and methodological approaches that enhance the reliability of these models' outputs should be developed.

# References

[1] J. Jiang, F. Wang, J. Shen, S. Kim, S. Kim, A Survey on Large Language Models for Code Generation, ACM Trans. Softw. Eng. Methodol. (2025). https://doi.org/10.1145/3747588.

[2] Y. Li, D. Choi, J. Chung, N. Kushman, J. Schrittwieser, R. Leblond, T. Eccles, J. Keeling, F. Gimeno, A. Dal Lago, Competition-level code generation with alphacode, Science (1979) 378 (2022) 1092–1097.

[3] C.S. Xia, Y. Wei, L. Zhang, Automated program repair in the era of large pre-trained language models, in: 2023 IEEE/ACM 45th International Conference on Software Engineering (ICSE), IEEE, 2023: pp. 1482–1494.

[4] S. Ouyang, J.M. Zhang, M. Harman, M. Wang, An empirical study of the non-determinism of chatgpt in code generation, ACM Transactions on Software Engineering and Methodology 34 (2025) 1–28.

[5] K. Liu, L. Li, A. Koyuncu, D. Kim, Z. Liu, J. Klein, T.F. Bissyandé, A critical review on the evaluation of automated program repair systems, Journal of Systems and Software 171 (2021) 110817.

[6] L. Gazzola, D. Micucci, L. Mariani, Automatic software repair: A survey, in: Proceedings of the 40th International Conference on Software Engineering, 2018: p. 1219.

[7] D. Lin, J. Koppel, A. Chen, A. Solar-Lezama, QuixBugs: A multi-lingual program repair benchmark set based on the Quixey Challenge, in: Proceedings Companion of the 2017 ACM SIGPLAN International Conference on Systems, Programming, Languages, and Applications: Software for Humanity, 2017: pp. 55–56.

[8] H. Ye, M. Martinez, T. Durieux, M. Monperrus, A comprehensive study of automatic program repair on the QuixBugs benchmark, Journal of Systems and Software 171 (2021) 110825.

[9] S. Ouyang, J.M. Zhang, M. Harman, M. Wang, An empirical study of the non-determinism of chatgpt in code generation, ACM Transactions on Software Engineering and Methodology 34 (2025) 1–28.

[10] S. Thakur, B. Ahmad, H. Pearce, B. Tan, B. Dolan-Gavitt, R. Karri, S. Garg, Verigen: A large language model for verilog code generation, ACM Transact Des Autom Electron Syst 29 (2024) 1–31.

[11] Y. Liu, C. Tantithamthavorn, Y. Liu, L. Li, On the reliability and explainability of language models for program generation, ACM Transactions on Software Engineering and Methodology 33 (2024) 1–26.

[12] Y. Liu, T. Le-Cong, R. Widyasari, C. Tantithamthavorn, L. Li, X.-B.D. Le, D. Lo, Refining chatgpt-generated code: Characterizing and mitigating code quality issues, ACM Transactions on Software Engineering and Methodology 33 (2024) 1–26.

[13] Y. Tang, Z. Liu, Z. Zhou, X. Luo, Chatgpt vs sbst: A comparative assessment of unit test suite generation, IEEE Transactions on Software Engineering 50 (2024) 1340–1359.

[14] Z. Liu, Y. Tang, X. Luo, Y. Zhou, L.F. Zhang, No need to lift a finger anymore? assessing the quality of code generation by chatgpt, IEEE Transactions on Software Engineering 50 (2024) 1548–1584.

[15] S. Kang, J. Yoon, N. Askarbekkyzy, S. Yoo, Evaluating diverse large language models for automatic and general bug reproduction, IEEE Transactions on Software Engineering (2024).

[16] Y. Yu, G. Rong, H. Shen, H. Zhang, D. Shao, M. Wang, Z. Wei, Y. Xu, J. Wang, Fine-tuning large language models to improve accuracy and comprehensibility of automated code review, ACM Transactions on Software Engineering and Methodology 34 (2024) 1–26.

[17] Y. Elazar, N. Kassner, S. Ravfogel, A. Ravichander, E. Hovy, H. Schütze, Y. Goldberg, Measuring and improving consistency in pretrained language models, Trans Assoc Comput Linguist 9 (2021) 1012–1031.

[18] L. Wang, X. Chen, X. Deng, H. Wen, M. You, W. Liu, Q. Li, J. Li, Prompt engineering in consistency and reliability with the evidence-based guideline for LLMs, NPJ Digit Med 7 (2024) 41.

[19] L. Fan, J. Liu, Z. Liu, D. Lo, X. Xia, S. Li, Exploring the Capabilities of LLMs for Code-Change-Related Tasks, ACM Transactions on Software Engineering and Methodology 34 (2025) 1–36.

[20] M. Mizrahi, G. Kaplan, D. Malkin, R. Dror, D. Shahaf, G. Stanovsky, State of what art? a call for multi-prompt llm evaluation, Trans Assoc Comput Linguist 12 (2024) 933–949.

[21] X. Zhou, S. Cao, X. Sun, D. Lo, Large language model for vulnerability detection and repair: literature review and the road ahead (2024), ArXiv Preprint ArXiv:2404.02525 (2024).

[22] J. Wang, Y. Huang, C. Chen, Z. Liu, S. Wang, Q. Wang, Software testing with large language models: Survey, landscape, and vision, IEEE Transactions on Software Engineering 50 (2024) 911–936.

[23] K. Tamberg, H. Bahsi, Harnessing Large Language Models for Software Vulnerability Detection: A Comprehensive Benchmarking Study. arXiv 2024, ArXiv Preprint ArXiv:2405.15614 (n.d.).

[24] N. Marques, R.R. Silva, J. Bernardino, Using ChatGPT in software requirements engineering: a comprehensive review. Future Internet 16 (6)(2024), (2024).

[25] X. Zhu, W. Zhou, Q.-L. Han, W. Ma, S. Wen, Y. Xiang, When software security meets large language models: A survey, IEEE/CAA Journal of Automatica Sinica 12 (2025) 317–334.

[26] X. Hou, Y. Zhao, Y. Liu, Z. Yang, K. Wang, L. Li, X. Luo, D. Lo, J. Grundy, H. Wang, Large language models for software engineering: A systematic literature review, ACM Transactions on Software Engineering and Methodology 33 (2024) 1–79.

[27] B. Ahmad, S. Thakur, B. Tan, R. Karri, H. Pearce, On hardware security bug code fixes by prompting large language models, IEEE Transactions on Information Forensics and Security 19 (2024) 4043–4057.

[28] Y. Dong, X. Jiang, Z. Jin, G. Li, Self-collaboration code generation via chatgpt, ACM Transactions on Software Engineering and Methodology 33 (2024) 1–38.

[29] M. Nashaat, J. Miller, Towards efficient fine-tuning of language models with organizational data for automated software review, IEEE Transactions on Software Engineering (2024).

[30] H. Li, Y. Hao, Y. Zhai, Z. Qian, Enhancing static analysis for practical bug detection: An llm-integrated approach, Proceedings of the ACM on Programming Languages 8 (2024) 474–499.

[31] L. Pan, M. Saxon, W. Xu, D. Nathani, X. Wang, W.Y. Wang, Automatically correcting large language models: Surveying the landscape of diverse automated correction strategies, Trans Assoc Comput Linguist 12 (2024) 484–506.

[32] M. Chelli, J. Descamps, V. Lavoué, C. Trojani, M. Azar, M. Deckert, J.-L. Raynier, G. Clowez, P. Boileau, C. Ruetsch-Chelli, Hallucination rates and reference accuracy of ChatGPT and bard for systematic reviews: comparative analysis, J Med Internet Res 26 (2024) e53164.

[33] Z. Lin, S. Guan, W. Zhang, H. Zhang, Y. Li, H. Zhang, Towards trustworthy LLMs: a review on debiasing and dehallucinating in large language models, Artif Intell Rev 57 (2024) 243.

[34] V.I. Levenshtein, Binary codes capable of correcting deletions, insertions, and reversals, Soviet Physics. Doklady 10 (1965) 707–710. https://api.semanticscholar.org/CorpusID:60827152.

[35] J. Li, G. Li, Z. Li, Z. Jin, X. Hu, K. Zhang, Z. Fu, Codeeditor: Learning to edit source code with pre-trained models, ACM Transactions on Software Engineering and Methodology 32 (2023) 1–22.

[36] J. Li, Y. Li, G. Li, Z. Jin, Y. Hao, X. Hu, Skcoder: A sketch-based approach for automatic code generation, in: 2023 IEEE/ACM 45th International Conference on Software Engineering (ICSE), IEEE, 2023: pp. 2124–2135.

[37] A. Mastropaolo, L. Pascarella, E. Guglielmi, M. Ciniselli, S. Scalabrino, R. Oliveto, G. Bavota, On the robustness of code generation techniques: An empirical study on github copilot, in: 2023 IEEE/ACM 45th International Conference on Software Engineering (ICSE), IEEE, 2023: pp. 2149–2160.

[38] D. Hendrycks, S. Basart, S. Kadavath, M. Mazeika, A. Arora, E. Guo, C. Burns, S. Puranik, H. He, D. Song, J. Steinhardt, Measuring Coding Challenge Competence With APPS, (2021). https://arxiv.org/abs/2105.09938.

[39] D. Zan, B. Chen, D. Yang, Z. Lin, M. Kim, B. Guan, Y. Wang, W. Chen, J.-G. Lou, CERT: Continual Pre-Training on Sketches for Library-Oriented Code Generation, (2022). https://arxiv.org/abs/2206.06888.

[40] D. Lin, J. Koppel, A. Chen, A. Solar-Lezama, QuixBugs: A multi-lingual program repair benchmark set based on the Quixey Challenge, in: Proceedings Companion of the 2017 ACM SIGPLAN International Conference on Systems, Programming, Languages, and Applications: Software for Humanity, 2017: pp. 55–56.